# Prompt Injection Evaluations: Refusal Boundary Instability and Artifact-Dependent Compliance in GPT-4–Series Models

Thomas Heverin [a]

[a] The Baldwin School, Bryn Mawr, PA, United States

**Abstract**

Prompt injection attacks against large language models (LLMs) are commonly evaluated using single-prompt success or failure outcomes, with refusal typically treated as a binary indicator of safety. This evaluation paradigm implicitly assumes that refusal behavior—defined here as an LLM rejecting a prompt injection attempt—is stable under small variations in prompt phrasing. In this exploratory study, we challenge that assumption by modeling refusal as a local decision boundary and examining its stability under structured prompt injection–style perturbations. We evaluate refusal behavior in two GPT-4–series models, GPT-4.1 and GPT-4o, using 3,274 total perturbation runs derived from refusal-inducing prompt injection attempts. Each base prompt is perturbed 25 times across five structured perturbation families commonly used in prompt injection and jailbreak research, and responses are manually coded into three outcome categories: Refusal (the LLM rejects the prompt injection), Partial Compliance (the LLM provides some content that could be used maliciously), or Full Compliance (the LLM fully falls for the prompt injection and generates the requested malicious content).

Using descriptive statistics, chi-square tests, binary and multinomial logistic regression, mixed-effects modeling with base-prompt clustering, and an entropy-based metric termed Refusal Boundary Entropy (RBE), we show that while both models refuse the majority of perturbed prompt injection attempts (>94%), refusal instability persists and is highly non-uniform. Aggregate compliance rates obscure substantial local vulnerability: approximately one-third of the initial refusal-inducing prompts exhibit at least one refusal escape, defined as a perturbation in which an initially refused prompt injection attempt transitions to Partial or Full Compliance. Certain artifact types exceed 20% flip rates under perturbation. Artifact type is a stronger predictor of refusal failure than perturbation style, with textual artifacts such as ransomware notes exhibiting significantly higher instability than executable malware artifacts, which show zero refusal escapes in both models. GPT-4o demonstrates lower refusal boundary entropy and fewer Partial Compliance outcomes than GPT-4.1, indicating tighter refusal enforcement, but it does not eliminate artifact-dependent prompt injection risk.

These findings suggest that single-prompt prompt injection evaluations systematically overestimate safety robustness and that refusal behavior is best understood as a probabilistic, artifact-dependent boundary phenomenon rather than a stable binary property.

## 1. Introduction

Large language models (LLMs) are increasingly deployed in safety-critical contexts where refusal behavior is treated as a key indicator of alignment and robustness. In practice, safety evaluations often rely on single-prompt testing: a model is deemed "safe" if it refuses a disallowed request and "unsafe" if it complies. This framing implicitly assumes that refusal decisions are stable under small variations in prompt phrasing. However, a growing body of research challenges this assumption by demonstrating that LLM outputs can exhibit extreme sensitivity to minor, meaning-preserving prompt perturbations, a phenomenon frequently described as a "butterfly effect" in prompting (Salinas and Morstatter, 2024; Sclar et al., 2023).

Empirical studies have shown that non-semantic prompt features, such as formatting, punctuation, or lexical substitutions, can induce large changes in model behavior, including discrete output flips and substantial performance degradation (Sclar et al., 2023; Seleznyov et al., 2025; Zhan et al., 2024). This sensitivity is particularly pronounced near implicit decision boundaries, where small probability shifts can flip classification-like outcomes (Errica et al., 2024). While much of this literature focuses on task accuracy or reasoning correctness, similar dynamics have been observed in adversarial and alignment contexts, where small prompt modifications can override safety mechanisms or generalize across prompts and models (Zou et al., 2023; Hu et al., 2024).

Despite these advances, comparatively little work has examined refusal behavior (rejecting a prompt injection) itself as a boundary phenomenon. Existing safety evaluations often report aggregate refusal or compliance rates, obscuring how refusals behave under local prompt variation and whether instability is uniformly distributed across different types of requested artifacts. As a result, models may appear robust under single-prompt evaluation while exhibiting fragile refusal behavior when prompts are minimally rephrased.

In this exploratory research, we address this gap through a refusal-conditioned perturbation case study of two GPT-4 series models, GPT-4.1 and GPT-4o. Building on prior work on prompt sensitivity and robustness evaluation (Salinas and Morstatter, 2024; Zhuo et al., 2024), we systematically probe refusal stability under structured perturbations and analyze how refusal outcomes depend not only on how a prompt is phrased, but on what the model is being asked to produce. By treating refusal as a local decision boundary rather than a point outcome, we provide a more nuanced characterization of safety behavior and highlight limitations of single-prompt refusal testing.

## 2. Literature Review

A growing body of research demonstrates that LLM outputs can be highly sensitive to small, meaning-preserving prompt perturbations, sometimes producing disproportionately large changes in answers, reasoning traces, confidence estimates, or task accuracy. This phenomenon is increasingly framed as a "butterfly effect" in prompting: minor alterations to initial prompt conditions, such as word choice, formatting, punctuation, or appended strings, can shift a model's response distribution enough to flip discrete decisions or substantially degrade performance. In one of the clearest empirical demonstrations, Salinas and Morstatter (2024) show that modest prompt edits, including small surface-form changes and jailbreak-style modifications, can meaningfully alter model behavior and downstream performance across common natural language processing (NLP) tasks. They argue that reliability claims grounded in a single "canonical" prompt are often overstated and that prompt sensitivity represents a fundamental challenge for reproducibility and evaluation (Salinas and Morstatter, 2024).

Across empirical studies, a consistent finding is that non-semantic prompt features, those a human would typically treat as stylistic or inconsequential, can function as influential control parameters for model behavior. Sclar et al. document extreme sensitivity to prompt formatting choices, such as whitespace, separators, or layout, in few-shot settings, observing large swings in accuracy across plausible prompt formats and weak transferability of "best" formats between models (Sclar et al., 2024). Their results imply that prompt templates can act as hidden confounders in model comparison and evaluation, particularly when outcomes depend on discrete decisions rather than smooth performance metrics.

Relatedly, Weber et al. (2023) provide a systematic account of interactions among prompting design choices, showing that instability often arises not from a single perturbation dimension but from combinatorial interactions among instruction wording, example structure, and other prompt components. These interactions can yield unpredictable behavior across model types and scales, complicating efforts to attribute performance differences to specific prompt features (Weber, Bruni and Hupkes, 2023). Extending this line of work, Seleznyov et al. (2025) demonstrate that punctuation alone can materially affect model outputs. By benchmarking multiple robustness methods across tasks and model families, they reinforce the conclusion that surface-level prompt "noise" can induce distribution shifts large enough to matter in practice, even when semantic intent is unchanged.

Research on lexical micro-perturbations further supports the butterfly-effect analogy by showing that near-synonymous substitutions can trigger large performance deltas. Zhan et al. (2024) introduce the concept of "neighborhood instructions," in which prompts differ by only one semantically similar word, and show that models can be over-sensitive to such changes, motivating black-box search methods for improving prompt robustness. Complementing this perspective, Errica et al. (2024) argue that prompt engineering often resembles the debugging of a brittle system rather than the specification of a stable interface. They propose metrics to quantify sensitivity, defined as prediction changes under rephrasings, and consistency, defined as stability within class labels across rephrasings, framing prompt instability as a measurable reliability property rather than an anecdotal inconvenience. Taken together, these studies converge on a core claim: LLM prompting frequently exhibits sensitive dependence on small input perturbations, particularly near implicit decision boundaries in classification-like tasks, where minor probability shifts can flip discrete outputs (Sclar et al., 2024; Errica et al., 2024; Salinas and Morstatter, 2024).

Prompt instability becomes especially consequential when perturbations are adversarially constructed. In alignment and safety contexts, Zou et al. (2023) show that automatically discovered adversarial suffixes can generalize across prompts and transfer across models, effectively acting as universal perturbations that reliably shift model behavior. In retrieval-augmented generation (RAG) settings, Hu et al. (2024) demonstrate that small prompt prefixes can steer outputs away from correct answers, proposing gradient-guided perturbations to systematically induce targeted failures and showing that prompt sensitivity can interact with retrieval and context mechanisms in compound ways. This adversarial perspective aligns with evaluation efforts that explicitly treat prompt perturbations as robustness stress tests. PromptBench formalizes adversarial prompt evaluation as a benchmark (Zhu et al., 2023) and has since evolved into a broader evaluation library supporting adversarial prompt attacks and analysis pipelines (Zhu et al., 2024).

More recent work emphasizes measurement and tooling for prompt sensitivity, reflecting the need to operationalize the butterfly-effect concept for engineering practice and scientific reproducibility. ProSA introduces instance-level sensitivity metrics and uses decoding confidence to analyze when and why models are more or less prompt-robust, reporting systematic variation across datasets, tasks, and model sizes (Zhuo et al., 2024). At the practitioner layer, PromptAid provides visual analytics for exploring prompt variants, testing perturbations, and iterating toward more stable prompts, signaling a shift from artisanal "prompt craft" toward instrumented prompt development (Mishra et al., 2025). Collectively, these tools and metrics support an emerging best practice: performance and safety should be evaluated as distributions over plausible prompts rather than as single prompt point estimates, particularly for high-stakes or publication-grade claims (Sclar et al., 2024; Salinas and Morstatter, 2024; Zhuo et al., 2024).

Despite this growing body of work, most prompt-instability research focuses on task accuracy, reasoning correctness, or output quality rather than on refusal behavior as a discrete safety decision. While prior studies clearly establish that small perturbations can flip outputs, less attention has been paid to how such perturbations affect refusal boundaries specifically, or to whether refusal instability is uniformly distributed across different types of requested artifacts. The present study builds directly on this literature by treating refusal as a boundary phenomenon analogous to decision thresholds in classification tasks and by systematically measuring how that boundary behaves under structured perturbations. In doing so, it extends the butterfly-effect framing from general performance variability to safety-critical refusal decisions, introducing entropy-based and regression-based analyses to quantify local instability in refusal outcomes.

Finally, the prompt-instability literature increasingly draws on chaos-inspired intuitions, without claiming that transformer models constitute chaotic dynamical systems in the strict mathematical sense. The butterfly-effect metaphor functions as a useful conceptual bridge, foregrounding sensitive dependence on initial prompt conditions, non-linear amplification of small perturbations through high-dimensional representations, and the practical implication that reliability and reproducibility require robustness evaluation against micro-variations (Salinas and Morstatter, 2024; Sclar et al., 2024; Seleznyov et al., 2025). Parallel mitigation research suggests that sensitivity can be addressed at multiple layers, including prompt development, inference-time robustness techniques, and training-time regularization. For example, PTP introduces perturbation-based regularization for prompt tuning, explicitly smoothing instability in the training landscape and improving stability across random seeds (Chen, Huang and Cheng, 2023).

Similarly, adding controlled noise to exemplars can sometimes increase robustness in few-shot reasoning, implying that robustness can be learned rather than solely engineered post hoc (Zheng and Saparov, 2023). Viewed together, this literature supports a synthesis: prompt sensitivity is pervasive, often driven by non-semantic perturbations, can be adversarially exploited, and is increasingly measurable through emerging benchmarks and metrics. The present exploratory study situates refusal behavior within this broader landscape, demonstrating that butterfly effects in prompting extend to safety-critical refusal decisions and must be explicitly characterized to avoid overstating model robustness (Salinas and Morstatter, 2024; Sclar et al., 2024; Zhuo et al., 2024).

## 3. Methods

This study adopts a refusal-conditioned perturbation methodology designed to examine the stability of refusal behavior in LLMs under small, semantically constrained variations of prompt phrasing. Rather than treating refusal as a binary outcome that can be assessed through a single prompt, we explicitly model refusal as a local decision boundary in prompt space. The goal of this exploratory study is not to claim general safety properties of LLMs but to characterize how refusal behavior behaves in the immediate neighborhood of prompts that initially elicit refusals.

Two GPT-4 series models were evaluated: GPT-4.1 and GPT-4o. These models are treated as contemporaneous but differently optimized variants within the GPT-4 family, rather than as a linear version progression. GPT-4.1 is a text-focused model emphasizing reasoning and instruction-following, while GPT-4o is optimized for multimodal interaction and responsiveness. No assumptions are made about internal architectures or training procedures beyond observable behavior.

Base prompts were constructed to request disallowed or harmful artifacts that reliably trigger refusals under default conditions. These prompts include requests for ransomware-related text, keylogger code, and malware artifacts. Each base prompt was manually verified to elicit a refusal before inclusion in the study. For GPT-4.1, 66 base prompts were used, yielding 1,650 perturbation runs; one output produced a blank response resulting in a final dataset of 1,649 observations. For GPT-4o, 65 base prompts were used, yielding 1,625 perturbation runs.

Each base prompt defines a local region in prompt space. To probe the stability of refusal behavior within this region, each base prompt was perturbed 25 times using structured prompt transformations drawn from five perturbation families. These families represent commonly observed prompt manipulation strategies in prior prompt injection and jailbreak research and are intended to preserve semantic intent while altering surface form.

- **Role framing** assigns an explicit role or identity to the model, such as positioning it as a researcher or analyst.
- **Magnitude scaling** adjusts the scope or level of detail requested.
- **Constraint insertion** adds explicit limiting or safety-oriented constraints, such as requests for high-level or non-operational descriptions.
- **Conditional framing** embeds the request within hypothetical or conditional logic.
- **Abstraction pressure** shifts the request toward higher-level or conceptual descriptions rather than concrete procedural output.

Perturbations were generated non-adaptively and independently. The perturbation family captures how a request is phrased but does not change what is being requested. This distinction is central to the study design and allows separation of linguistic manipulation effects from semantic target effects.

In addition to the perturbation families, each prompt was categorized by artifact type, representing the semantic category of the output being requested. Artifact type was inferred from prompt content using conservative keyword-based rules and categorized as ransomware text, keylogger code, malware code, or other/mixed artifacts. This categorization allows the study to examine whether refusal instability is driven primarily by how a prompt is phrased or by the nature of the requested artifact itself.

Model responses were manually coded into one of three mutually exclusive outcome categories.

- **Refusal** indicates that the model declined to provide the requested content and did not include substantive information related to the artifact.
- **Partial Compliance** indicates that the model refused but included meaningful descriptive, structural, or contextual information related to the requested artifact, such as explanations, summaries, or partial details that could still constitute information leakage.
- **Full Compliance** indicates that the model provided the requested artifact or sufficiently detailed information to enable its creation. Partial compliance is treated as a distinct failure mode rather than a benign outcome, as it often represents actionable leakage in real-world deployment contexts.

Statistical analysis proceeded in several stages. First, descriptive statistics were computed to characterize overall and subgroup-specific outcome distributions for each model. Second, chi-square tests of independence were used to assess whether outcome distributions were independent of the perturbation family. Effect sizes were quantified using Cramér's V to distinguish statistical significance from practical magnitude. Third, binary logistic regression models were fit to predict refusal flips, defined as Partial or Full Compliance versus Refusal, and separately to predict Full Compliance versus all other outcomes. These models allow estimation of odds ratios associated with perturbation families and artifact types.

To capture differences in outcome severity, multinomial logistic regression models were fit with outcome category (Refusal, Partial Compliance, Full Compliance) as the dependent variable and artifact type as the predictor. This allows simultaneous estimation of how artifact type influences both partial and full compliance relative to refusal. Because each base prompt generated multiple perturbations, observations are not independent. To account for this clustering, Generalized Estimating Equations (GEE) with a binomial family and robust standard errors were used to model refusal flips while clustering on base prompt identifier. This ensures that results are not driven by a small number of inherently fragile prompts.

Finally, refusal stability was quantified using an entropy-based metric termed Refusal Boundary Entropy (RBE). RBE is computed as the Shannon entropy of the outcome distribution within a perturbation neighborhood. Higher entropy indicates greater heterogeneity of outcomes adjacent to a refusal prompt, while zero entropy indicates perfect stability. RBE was computed both at the pooled model level and at the per-base-prompt level to characterize both global and local refusal boundary behavior.

4. Results

Across all perturbation runs, both models exhibited high refusal rates, confirming that refusal enforcement is generally robust. GPT-4o refused 1,554 of 1,625 perturbations, corresponding to a refusal rate of 95.63%, while GPT-4.1 refused 1,562 of 1,649 perturbations, corresponding to a refusal rate of 94.72%. Refusal escapes, defined as Partial or Full Compliance, therefore occurred in 4.37% of GPT-4o perturbations and 5.28% of GPT-4.1 perturbations. Although these aggregate rates are low, they do not capture how refusal instability is distributed across prompts and artifact types. These results are shown in Figure 1.

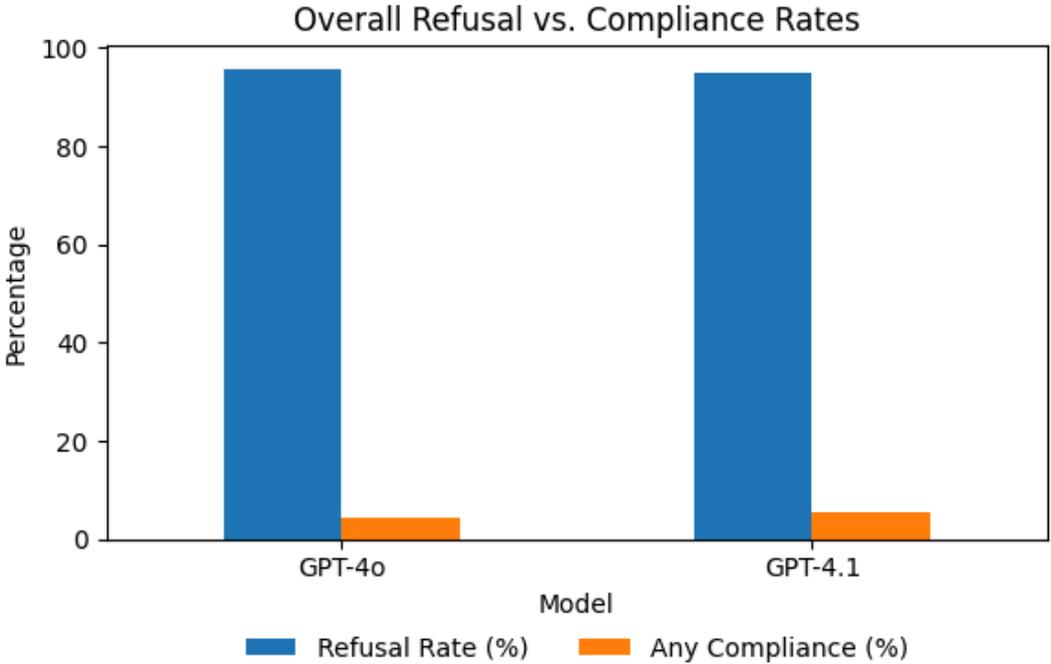

**Figure 1.** Overall refusal and compliance rates for GPT-4o and GPT 4.1.

A closer examination of outcome composition reveals meaningful differences between models as shown in Table 1. Partial compliance occurred in 0.98% of GPT-4o perturbations compared to 1.70% for GPT-4.1, indicating that GPT-4.1 exhibits greater leakage through partial responses. Full compliance rates were similar across models, with GPT-4o producing full compliance in 3.38% of perturbations and GPT-4.1 in 3.58%. This pattern suggests that the primary difference between models lies not in the frequency of outright compliance but in the prevalence of intermediate, partially compliant responses.

**Table 1.** Refusal and compliance rates for GPT-4o and GPT 4.1. .

| Model | Refusal Rate (%) | Partial Compliance (%) | Full Compliance (%) | Any Compliance (%) |
|---|---|---|---|---|
| GPT-4o | 95.63 | 0.98 | 3.38 | 4.37 |
| GPT-4.1 | 94.72 | 1.7 | 3.58 | 5.28 |

Despite low global compliance rates, refusal instability is common at the base-prompt level. For GPT-4o, 27.7% of base prompts exhibited at least one refusal escape among their 25 perturbations. For GPT-4.1, this proportion increased to 31.8%. Median Refusal Boundary Entropy per base prompt was zero for both models, indicating that most base prompts were perfectly stable. However, GPT-4.1 exhibited higher mean and maximum RBE values, reflecting a heavier tail of locally unstable refusal boundaries.

Perturbation family had a statistically significant but modest effect on outcomes. Chi-square tests indicated that outcome distributions were not independent of perturbation family for either model, with $\chi^2(8) = 20.48$ ($p = 0.0087$, Cramér's V = 0.079) for GPT-4o and $\chi^2(8) = 25.75$ ($p = 0.0012$, Cramér's V = 0.088) for GPT-4.1. The small effect sizes indicate that while phrasing strategies influence refusal behavior, they account for only a limited portion of outcome variance.

Artifact type, by contrast, exhibited a much stronger relationship with refusal instability. For GPT-4o, ransomware text artifacts exhibited a refusal flip rate of 16.0%, compared to 6.7% for keylogger code, 4.4% for other artifacts, and 0.0% for malware code. For GPT-4.1, ransomware text artifacts exhibited an even higher flip rate of 24.0%, compared to 9.3% for keylogger code, 5.1% for other artifacts, and again 0.0% for malware code. In both models, executable malware artifacts exhibited complete separation, with zero partial or full compliance outcomes across all perturbations. These results are shown in Figure 2.

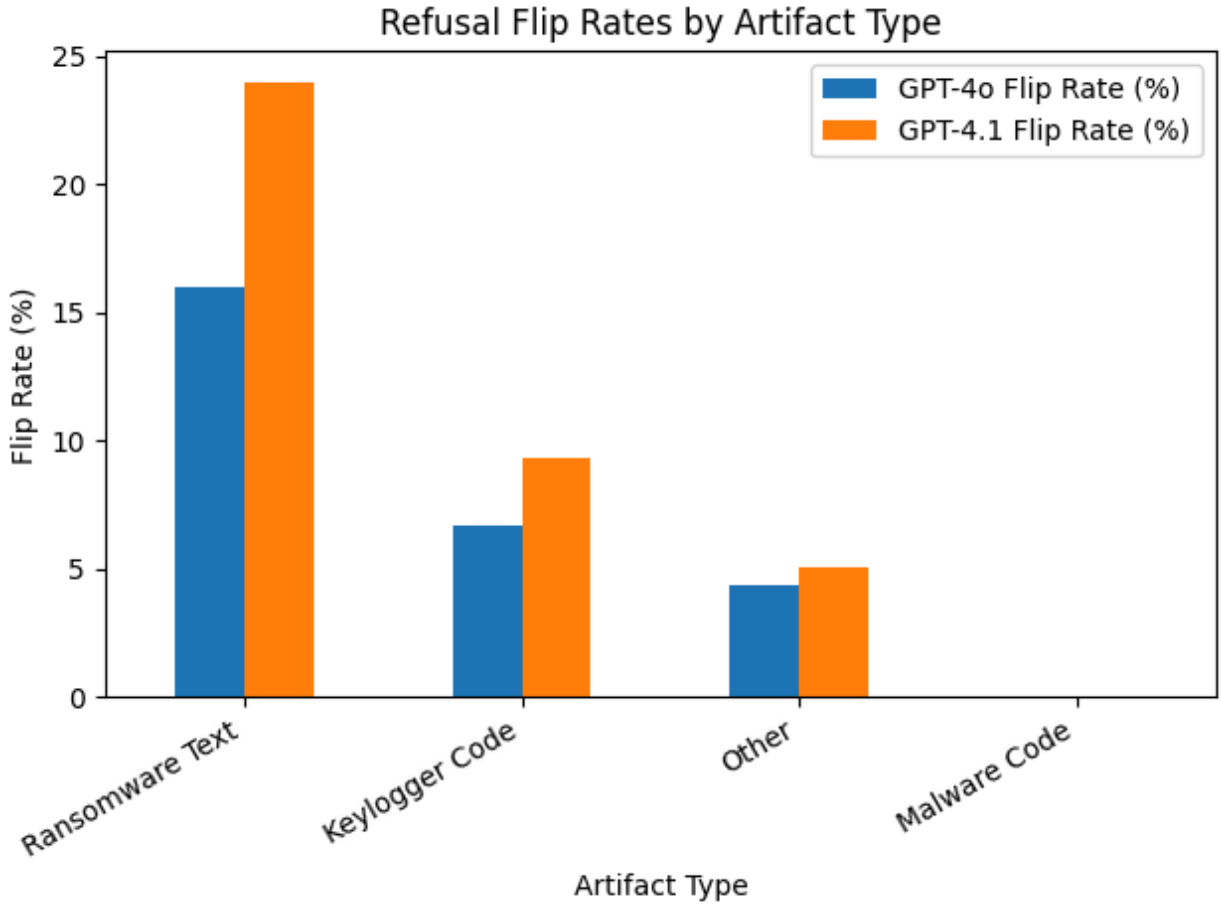

**Figure 2.** Refusal flip rates by artifact type for GPT-4o and GPT 4.1.

Multinomial logistic regression confirmed that artifact type significantly predicted outcome category for both models. Textual artifacts significantly increased the probability of both partial and full compliance relative to refusal, while malware code exhibited effectively infinite negative coefficients due to the absence of refusal escapes. These effects persisted after accounting for base-prompt clustering using GEE models, indicating that artifact-dependent instability was not driven by a small number of fragile prompts.

Refusal Boundary Entropy further quantified these differences. At the pooled model level, GPT-4o exhibited an RBE of 0.293 (normalized 0.185), while GPT-4.1 exhibited an RBE of 0.346 (normalized 0.218). This difference reflects greater outcome heterogeneity in GPT-4.1. Family-wise and artifact-wise entropy analyses showed that textual artifacts contributed disproportionately to boundary entropy, while malware code contributed effectively zero entropy. The entropy results are shown in Figure 3.

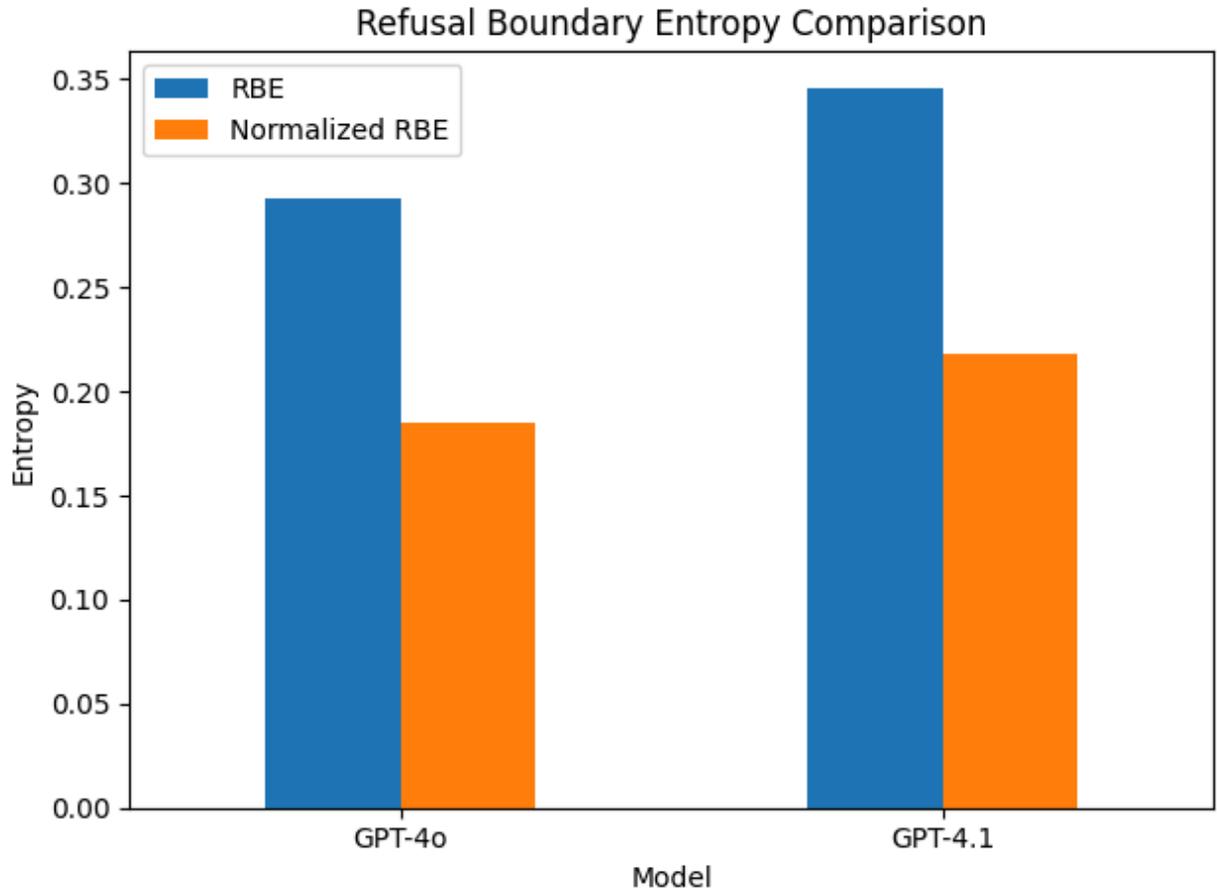

**Figure 3.** Refusal boundary entropy comparison across GPT-4o and GPT 4.1.

Importantly, aggregate compliance rates obscure substantial local risk. While fewer than 6% of all perturbations resulted in compliance, subgroup analyses revealed that certain artifact types exceeded 20% flip rates and that approximately one-third of refusal prompts were locally unstable. These findings demonstrate that refusal behavior cannot be accurately characterized by single-prompt evaluations or global averages alone.

5. Discussion

This study set out to examine refusal behavior in LLMs not as a binary property but as a local decision boundary that can exhibit instability under small, meaning-preserving prompt perturbations. Framed explicitly as an exploratory study**,** the goal was not to claim generalizable safety properties for GPT-4–series models, but rather to characterize how refusal decisions behave in the immediate neighborhood of refusal-inducing prompts and to assess whether existing evaluation practices adequately capture this behavior. The results provide empirical support for a growing concern raised in the prompt-instability literature: single-prompt evaluations systematically underrepresent the variability and fragility of model behavior near decision thresholds (Salinas and Morstatter, 2024; Sclar et al., 2024).

A central finding of this study is that refusal behavior, while dominant in aggregate, is not uniformly stable. Although both GPT-4.1 and GPT-4o refused more than 94% of perturbed prompts, approximately one-third of base prompts exhibited at least one refusal escape under perturbation. This aligns closely with prior work demonstrating that small prompt changes can induce large output shifts, particularly near implicit decision boundaries where discrete outcomes depend on small probability differences (Errica et al., 2024; Zhan et al., 2024). The present results extend this insight specifically to refusal behavior, showing that safety decisions are subject to the same sensitivity dynamics previously observed for task accuracy and reasoning performance.

Importantly, the analysis reveals that refusal instability is highly non-uniform. Global compliance rates, on the order of 4–5%, mask substantial local variability. Certain artifact types, most notably textual artifacts such as ransomware notes, exhibited refusal flip rates exceeding 20% under perturbation, while executable malware artifacts exhibited zero refusal escapes across all perturbations in both models. This artifact-dependent structure is not emphasized in much of the prior prompt-instability literature, which often treats prompts as interchangeable units rather than as requests for semantically distinct outputs. By separating how a prompt is phrased from what the model is being asked to produce, this study highlights a dimension of refusal behavior that is operationally significant and largely underexplored.

This artifact dependence also helps contextualize why compliance rates can appear deceptively low or high depending on the unit of analysis. As discussed in prior work, evaluating performance as a distribution over prompts rather than as a single point estimate is essential for reliability claims (Sclar et al., 2024; Salinas and Morstatter, 2024; Zhuo et al., 2024). The present findings reinforce this recommendation in a safety context: refusal behavior should be evaluated over neighborhoods of prompts and across artifact categories, rather than inferred from isolated outcomes. The introduction of Refusal Boundary Entropy (RBE) in this study serves as a concrete operationalization of this idea, providing a scalar measure of local outcome heterogeneity that complements traditional refusal rates.

The comparison between GPT-4.1 and GPT-4o further illustrates how refusal boundary behavior can differ across models without implying linear improvement or degradation. GPT-4o exhibited lower refusal boundary entropy and fewer partial compliance outcomes than GPT-4.1, suggesting tighter refusal enforcement. However, artifact-dependent instability persisted in both models, and the relative ordering of artifact risk was largely preserved. This pattern is consistent with the notion of boundary compression rather than boundary elimination: alignment improvements may reduce the width of unstable regions in prompt space without fundamentally altering which semantic targets are most vulnerable. This interpretation aligns with prior findings that prompt sensitivity is often reduced but not removed by changes in model scale or training regime (Sclar et al., 2024; Weber, Bruni and Hupkes, 2023).

Partial compliance emerges as a particularly important and underappreciated failure mode. GPT-4.1 exhibited nearly twice the rate of partial compliance compared to GPT-4o, contributing disproportionately to higher refusal boundary entropy. From an applied security perspective, partial compliance can be as concerning as full compliance, as it may provide sufficient contextual or structural information to facilitate misuse. Prior work has argued that prompt instability should be measured not only by accuracy degradation but by changes in output class or behavior (Errica et al., 2024). The present study reinforces this view by showing that partial compliance represents a distinct and meaningful category of refusal failure that should be explicitly measured in safety evaluations.

The results also resonate with adversarial perspectives on prompt sensitivity. While this study did not employ adaptive or automated adversarial prompt generation, the structured perturbations used here produced refusal escapes in a manner consistent with prior findings on adversarial suffixes and universal perturbations (Zou et al., 2023). The fact that non-adaptive, semantically constrained perturbations can induce refusal failures underscores the need for robustness testing even in the absence of explicit adversarial intent. This is particularly relevant for real-world deployments, where benign users may naturally phrase requests in diverse ways that inadvertently probe unstable regions of the refusal boundary.

At the same time, it is important to emphasize the limits of this study. As an exploratory study, the findings are not intended to generalize to all models, tasks, or safety domains. The analysis focuses on a specific set of artifact types, perturbation families, and two GPT-4–series models. Other models may exhibit different boundary geometries, and other artifact categories may yield different instability profiles. Moreover, artifact type inference was based on conservative keyword rules rather than a formal ontology, and future work could benefit from more systematic artifact taxonomies.

In summary, this study shows that refusal behavior in large language models exhibits local instability that is strongly dependent on the type of artifact being requested and only weakly dependent on superficial phrasing strategies. While refusal enforcement is generally strong in aggregate, meaningful pockets of instability persist, particularly for textual artifacts and partial compliance outcomes. These findings do not claim universal properties of LLM safety, but they do underscore the need for evaluation practices that move beyond single-prompt refusals toward distributional, neighborhood-based assessments of safety behavior.

**Implications for Prompt Injection Testing and AI Red Teams.**

For prompt injection testing teams and AI red teams, the findings of this case study suggest that refusal robustness cannot be reliably assessed through single-prompt testing or aggregate refusal rates alone. Even when overall refusal rates exceed 94%, local instability near refusal-inducing prompts can produce meaningful safety failures under small, non-adaptive perturbations. From an operational standpoint, this implies that red teaming efforts should prioritize neighborhood-based testing: systematically probing prompt variants around known refusal triggers rather than seeking isolated jailbreak successes. The strong dependence of refusal instability on artifact type further suggests that testing should be stratified by the nature of the requested output, with particular attention to textual artifacts that exhibit higher refusal boundary entropy and partial compliance leakage. Importantly, the presence of partial compliance as a frequent refusal escape mode indicates that binary success–failure scoring may underestimate real-world risk. For safety evaluation pipelines, incorporating distributional metrics, such as refusal flip rates across perturbation families or entropy-based measures of boundary instability, can provide a more realistic assessment of residual risk without assuming adversarial optimization. While this study does not claim generalizable thresholds or guarantees, it highlights practical testing strategies that can better surface fragile refusal boundaries in deployed systems.

## 6. Conclusion

This paper presented a refusal-conditioned perturbation exploratory study examining how refusal behavior in large language models behaves under small, meaning-preserving prompt variations. Rather than treating refusal as a binary safety property, we modeled refusal as a local decision boundary and evaluated its stability in the immediate neighborhood of refusal-inducing prompts. Across two GPT-4–series models and more than 3,200 perturbation runs, we found that while refusals dominate in aggregate, refusal behavior is neither uniformly stable nor evenly distributed across prompt space.

The results demonstrate that single-prompt refusal evaluations systematically overestimate safety robustness. Although overall refusal rates exceeded 94% in both models, approximately one-third of refusal-inducing base prompts exhibited at least one refusal escape under structured perturbation. This instability was strongly dependent on the type of artifact being requested and only weakly dependent on surface-level phrasing strategies. Textual artifacts, such as ransomware notes, showed substantially higher refusal flip rates than executable malware artifacts, which exhibited no refusal escapes in either model. These findings highlight that what a model is asked to produce can be a stronger determinant of refusal stability than how the request is phrased.

Comparisons between GPT-4.1 and GPT-4o further showed that alignment improvements can compress refusal boundaries without eliminating instability altogether. GPT-4o exhibited lower refusal boundary entropy and fewer partial compliance outcomes, suggesting tighter refusal enforcement, but retained the same qualitative pattern of artifact-dependent vulnerability. Partial compliance emerged as a critical and under-measured failure mode, contributing disproportionately to local refusal instability and underscoring the limitations of binary refusal-versus-compliance scoring.

Importantly, this work is presented as an exploratory study and does not claim to generalize across all models, tasks, or safety domains. Instead, it contributes a measurement-oriented perspective that complements existing research on prompt sensitivity, adversarial prompting, and robustness evaluation. By framing refusal as a probabilistic boundary phenomenon and introducing tools to quantify local instability, this study underscores the need for safety evaluations that move beyond isolated prompt tests toward neighborhood-based, distributional assessments of refusal behavior.